\documentclass[12pt]{article}

\usepackage[english]{babel}

\begin{document}


\vskip 7mm

\begin{center}
{\Large \sc EXPONENTIAL WEALTH DISTRIBUTION IN DIFFERENT DISCRETE ECONOMIC MODELS} \\

\vskip 4mm
{\sc L\'opez-Ruiz  ~RICARDO}\\

\vskip 1mm
{\it Universidad de Zaragoza, \\
 Zaragoza, Spain} \\
{\small e-mail: rilopez@unizar.es}  
\end{center}

\bigskip

Exponential distribution is ubiquitous in the framework of multi-agent systems. 
Usually, it appears as an equilibrium state in the asymptotic time evolution of
statistical systems. It has been explained from very different perspectives.
In statistical physics, it is obtained from the principle of maximum entropy \cite{jaynes1957}.
In the same context, it can also be derived without any consideration about information theory,
only from geometrical arguments under  the hypothesis of equiprobability in phase space \cite{lopez2008}. 
Also, several multi-agent economic models based on mappings, with random, deterministic or 
chaotic interactions, can give rise to the asymptotic appearance of the exponential wealth distribution 
\cite{yako2001,sanchez2007,pellicer2010}. An alternative approach to this problem
in the framework of iterations in the space of distributions will be presented \cite{calbet2010}.
Concretely, the new iteration given by
\begin{equation}
f_{n+1}(x) = \int\!\!\int_{u+v>x}\,{f_n(u)f_n(v)\over u+v} \; dudv \,.
\label{syst1}
\end{equation}
\noindent It is found that the exponential distribution is a stable fixed point of
this type of systems (\ref{syst1}). From this point of view, it is easily understood 
why the exponential wealth distribution (or by extension, other kind of distributions) 
is asymptotically obtained in different multi-agent economic models (*).

\noindent
{\scriptsize (*) Communication presented in ECIT 2010,\\
{\it \scriptsize European Conference on Iteration Theory}, \\
{\scriptsize September 12 - 17, 2010, ~Nant, ~France}.

\end{document}